\documentclass[epj]{svjour}

\usepackage{latexsym}
\usepackage{graphics}
\usepackage{hyperref}
\usepackage{indentfirst}
\usepackage{bm}
\usepackage{cite} 
 \hypersetup{colorlinks=true, urlcolor=blue, 
linkcolor=blue, citecolor=blue} 
\usepackage{lineno,hyperref}
\usepackage{graphicx}
\usepackage{amsmath}
\usepackage{amssymb}
\usepackage{mathrsfs}
\usepackage{tikz}
\usepackage{filecontents}
\begin{document}
\title{Entanglement in a fermionic spin chain containing a single mobile boson under decoherence}
\author{Hamid Arian Zad \inst{1} 
\thanks{\emph{}
e-mail: arianzad.hamid@mshdiau.ac.ir}
 \and Majid Moradi (MKMajid) \inst{2}
 \thanks{\emph{}
e-mail: majidofficial@gmail.com}
}

%
%
\institute{Young Researchers and Elite Club, Mashhad Branch, Islamic Azad University, Mashhad, Iran
\and Physics Department, Shahrood University of Technology, Shahrood, Iran}

%
%
\abstract{
 The concurrence between first and the last sites of a fermionic spin chain containing a single boson is rigorously investigated at finite low temperature in the vicinity of a weak homogeneous magnetic field. We consider the boson as a mobile spin-1 particle through the chain and study concurrence without/under decoherence and express some interesting phase flip and bit flip reactions of the pairwise entanglement between first and the last half-spins in the chain. Our investigations show that the concurrence between two considered half-spins has different behavior for various positions of the single boson along the chain. Indeed, we realize that the single boson mobility has an essential role to probe the pairwise entanglement intensity between two spins located at the opposite ends of a fermionic chain. Interestingly, the entanglement remains alive for higher temperatures when the boson is the nearest neighbor of the first fermion. When the boson is at the middle of chain, it is demonstrated that the threshold temperature (at which the concurrence vanishes) versus decoherence rate can be considered as a threshold temperature boundary. These results pave the way to set and interpret the numerical and analytical expressions for utilizing quantum information in realistic scenarios such as quantum state transmission, quantum communication science and quantum information processing, where the both  fermion-fermion and fermion-boson correlations should be taken in to account.}
%
%
\maketitle
\section{Introduction}
Up to now, there was a tremendous interest for investigating entanglement of various Heisenberg models
 \cite{ABGARYAN20155,Sengupta2009,Horodecki2009,Ananikian2012,Wootters1998,Langari2008,TWelang}.There have been a great number of studies focusing on the applications of entanglement in quantum teleportation, spintronics and quantum information processing \cite{Mcmahon2007,Barmettler2010,Dolde2013,Furman2012}. Spin models are known as simple popular quantum systems in solid state physics and quantum comuting which exhibits remarkable entanglement features  \cite{Ollivier2001,Canosa2004,Zhang2005,GHAHARI2007,ZHANG2011}, specially quantum antiferromagnetic spin chains have been the subject of numerous experimental as well as theoretical studies for a long period of time \cite{Tonegawa1998}. A wide variety of the analytical and numerical techniques have been developed in order to study Ising-Heisenberg spin models \cite{ABGARYAN20155,Torrico2014,Chuang2008,Zad2017,Rojas2012,ABGARYAN201510} due to enhance the applications of quantum information theory in solid state physics and condensed matter such as investigating thermal entanglement for the mixed spin chains \cite{Tonegawa1998,Durga2013,Movahedian2017,Arian2018,Arian2018S,Zhe2009}.
 
  Motivated by the real materials it has already been proposed the class of one-dimensional plaquette chains to investigate the magnetic properties as well as quantum state transferring  from theoretical and analytical points of view \cite{Miyahara1999,Souza2016}.  Besides the frustrated spin-$S$ orthogonal-dimer chain, Koga {\it et al.} investigated in detail the first-order quantum phase transition of the orthogonal-dimer spin chain \cite{Koga2000,Koga2002}.
  A wide variety of the orthogonal dimer plaquette investigations have been carried out from theoretical and experimental view points. Driven by the comments given in Ref. \cite{Paulinelli2013}, we here consider the orthogonal dimer plaquette chain with antiferromagnetic coupling as schematically described in Fig. \ref{Fig-CentBoson} with an analytical Hamiltonian. On the other hand, the spin-impurity atoms were also widely discussed by some authors, for instance, T. Fukuhara {\it et al.} \cite{Fukuhara2013} studied the quantum dynamics created by the spin-impurity atom, as it propagates in a one-dimensional spin chain.  A specific probability distribution of the impurity at different times using single-site-resolved imaging of bosonic atoms  have been probed as well. Their experimental results open new prospects to studying mobile impurities in quantum spin systems specially quantum fluids. Accordingly, we consider the dimer plaquette as a single mobile boson in a one-dimensional spin chain which dramatically affects on the thermal entanglement between a pair of fermions. 

Heretofore, the dynamic behavior of spin chains under decoherence has been particularly studied \cite{Li2011,Maziero2009,Zad2016}. Maziero {\it et.al.}  investigated rigorously the dynamics of quantum and classical correlations under decoherence and carried out some interesting outcomes: classical correlation remains constant over time and quantum correlation decays monotonically over time \cite{Maziero2009}. During last decade, similar tasks has been made to understand defect production due to the dynamics of a spin model through a critical point \cite{Sengupta2009,Zurek2015,Sengupta2008}.

Because quantum communications allows to have unconditionally stable communication, and on the other hand, a quantum processor could dramatically speed up computations, there is a great deal of attentions in techniques to protect quantum information from noise \cite{Smith1812}. Hence, in this paper we are going to study and compare the entanglement of three different types of antiferromagnetic fermionic chains with defect caused by a boson under decoherence. In this regard, we consider the entanglement between first and the last half-spins in a finite size fermionic chain consisting of a single mobile boson. Three types of the suggested model with different spin arrangement are determined by changing the position of the single boson along the chain.

The paper is organized as follows. In Sec. \ref{Model} we first introduce the three sub-models with analytical Hamiltonians and describe the concurrence as a measure of entanglement by characterizing pairwise reduced density matrix. In Sec. \ref{decoherency} we show the numerical calculations and simulations of the pairwise entanglement under decoherence.  Section \ref{Conclusions} is devoted to conclusions.

\section{The Model}\label{Model}
In the present work, we assume an antiferromagnetic spin-1/2 chain with the total number of sites $N=5$ containing a single boson and four fermions with three different types of the spin arrangement, namely, we consider various locations for the boson along the chain and create three dissimilar sub-models. We witness that different positions of the boson along the chain introduce different properties for the entanglement between two fermions located at the opposite ends of the chain. To study the behavior of such systems, their Hamiltonians should be introduced. Firstly, we start our investigations by contemplating the boson at the middle of chain. So, the Hamiltonian can be expressed as 
\begin{equation}\label{hamiltonian}
\begin{array}{lcl}
H=\sum\limits_{i=1}^{\frac{N-3}{2}} \big[ J_{1}\big( {\sigma}^{x}_{i}{\sigma}^{x}_{i+1}+{\sigma}^{y}_{i}{\sigma}^{y}_{i+1}\big) + \Delta_{1}  {\sigma}^{z}_{i}{\sigma}^{z}_{i+1}\big]+  \\
J_{2}\big({\sigma}_{\frac{N-1}{2}}^{x}{J}^{x}+{J}^{x}{\sigma}_{\frac{N+3}{2}}^{x}+{\sigma}_{\frac{N-1}{2}}^{y}{J}^{y}+{J}^{y}{\sigma}_{\frac{N+3}{2}}^{y}\big)
\\
+\Delta_{2}\big({\sigma}_{\frac{N-1}{2}}^{z}{J}^{z}+{J}^{z}{\sigma}_{\frac{N+3}{2}}^{z}\big) +\\
\sum\limits_{i=\frac{N+3}{2}}^{N-1} \big[ J_{1}\big( {\sigma}^{x}_{i}{\sigma}^{x}_{i+1}+{\sigma}^{y}_{i}{\sigma}^{y}_{i+1}\big) + \Delta_{1}  {\sigma}^{z}_{i}{\sigma}^{z}_{i+1}\big] + \\
+B_{z} \cdot \big(J^{z} + \sum\limits_{i=1}^{N} \sigma_{i}^{z} \big),
\end{array}
\end{equation}
where $N$ is the total number of particles (here $N=5$), $J_{1}$ denotes exchange interaction between half-spins, $J_{2}$ is  the exchange interaction between the boson and its nearest fermions. $\Delta_{1}$ and $\Delta_{2}$ are anisotropy properties between, respectively,  fermion-fermion and fermion-boson. $B_{z}$ is the homogeneous magnetic field applied along the z-axis. Here,  ${\boldsymbol\sigma}=\lbrace {\sigma}^x, {\sigma}^y, {\sigma}^z \rbrace$ are the Pauli operators and
\begin{equation}
\begin{array}{lcl}
{J^x} = \left(
\begin{array}{ccc}
0 & 1 & 0\\
1 & 0 & 1\\
0 & 1 & 0 \\
\end{array} \right), 
 {J^y} = \left(
\begin{array}{ccc}
0 & -i & 0\\
i & 0 & -i\\
0 & i & 0 \\
\end{array} \right),
{J^z} =\left(
\begin{array}{ccc}
1 & 0 & 0\\
0 & 0 & 0\\
0 & 0 & -1 \\
\end{array} \right).
\end{array}
\end{equation}

After driving eigenvalues and eigenvectors of the Hamiltonian (\ref{hamiltonian}), in the standard basis, the total density matrix ${\rho}$ in equilibrium can be given by $\rho_{eq}=\frac{1}{\mathcal{Z}}\exp (-\beta H)$.
Regarding this, we can extract reduced density matrix of the half-spins at two opposite ends of the chain as
\begin{equation}\label{density matrices}
{{\rho}_{1N}} =\frac{1}{\mathcal{Z}}\left(
\begin{array}{cccc}
\alpha_{11} & \alpha_{12} & \alpha_{13} & \alpha_{14}\\
\alpha_{21} & \alpha_{22} & \alpha_{23} & \alpha_{24} \\
\alpha_{31} & \alpha_{32} & \alpha_{33} & \alpha_{34}\\
\alpha_{41} & \alpha_{42} & \alpha_{43} & \alpha_{44}
\end{array} \right),
\end{equation}
where $\mathcal{Z}=Tr[\exp(-\beta H)]$ is thermal partition function of the total system. $\beta=1/k_BT$ (we set $k_B=1$) and $T$ is the temperature.

In order to study thermal pairwise entanglement, we use the concurrence which can be defined by \\
$C(\tilde{\rho})=\max\{0,2\lambda-\sum^{4}_{i=1}\lambda_i\}$, where $\lambda=\max\{\lambda_1,\lambda_2,\lambda_3,\lambda_4\}$ and $ \lambda_i$  are square roots of the eigenvalues of the inner product $R=\rho . \tilde{\rho}$ with $\tilde{\rho}=(\sigma_y\otimes\sigma_y){\rho^\dagger}(\sigma_y\otimes\sigma_y)$  ($\rho^{\dagger}$ is the complex conjugate of matrix $\rho$).

\section{decoherency}\label{decoherency}
Recently, it has been revealed that the concurrence of bipartite subsystem (1/2,1/2) in a tripartite mixed spin chain (1/2,1,1/2) undergoing only the phase flip channel by means of the Kraus operators, is different in behavior \cite{Zad2016}. In this work, we investigate the thermal entanglement  for two half-spins at two opposite ends of the spin chain including a single boson under decoherence by means of the concurrence. Here, the state of bipartite system introduced by (\ref{density matrices}) is exactly investigated when it undergoes both phase flip and bit flip.

\subsection{Phase flip channel.} 
Kraus operators for the phase flip channel are given by $\Gamma_{{A}}^0=diag(\sqrt{1-p/2},\sqrt{1-p/2})\otimes I_{B}$, 
$\Gamma_{{A}}^1=diag(\sqrt{p/2},-\sqrt{p/2})\otimes I_{B}$,
 $\Gamma_{{B}}^0=I_{A}\otimes diag(\sqrt{1-p/2},\sqrt{1-p/2})$ and $\Gamma_{{B}}^1=I_{A} \otimes diag(\sqrt{p/2},-\sqrt{p/2})$ that describe the noise channels ${A}$ and ${B}$, where $0\leq p\leq1$. The evolved state under the phase flip channel can be characterized by a completely positive trace preserving (CPTP) map \\
 $\mathcal{\xi}\big[\rho(A B)\big]=\sum\limits_{i,j}\Gamma_{{A}}^{i}\Gamma_{{B}}^{j}\rho ({A}{B})\Gamma_{{B}}^{i\dagger}\Gamma_{{A}}^{j\dagger}$, where $A$ is the first site of the chain under consideration and $B$ is its last half-spin.

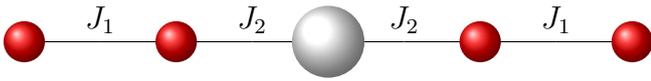
\begin{figure}[h!]
	\begin{center}
		\begin{tikzpicture}
		\draw[-] (0,0) -- (1,0) node[above] {\large{$J_{1}$}};
		\draw[-] (1,0) -- (2,0);
		\draw[-] (2,0) -- (3,0) node[above] {\large{$J_{2}$}};
		\draw[-] (3,0) -- (4,0);
		\draw[-] (4,0) -- (5,0) node[above] {\large{$J_{2}$}};
		\draw[-] (5,0) -- (6,0);
		\draw[-] (6,0) -- (7,0) node[above] {\large{$J_{1}$}};
		\draw[-] (7,0) -- (8,0);
		\shade[ball color=red] (0,0) circle (2ex);
		\shade[ball color=red] (2,0) circle (2ex);
		\shade[ball color=white] (4,0) circle (3.5ex);
		\shade[ball color=red] (6,0) circle (2ex);
		\shade[ball color=red] (8,0) circle (2ex);
		\end{tikzpicture}
	   \caption{Schematic representation of the spin chain containing a single boson located in the middle with spin arrangement (1/2,1/2,1,1/2,1/2)(sub-model 1).}
	    \label{Fig-CentBoson}
	\end{center} 
\end{figure}

\begin{figure}
	\begin{center}
		\includegraphics[width=7cm,height=5.5cm]{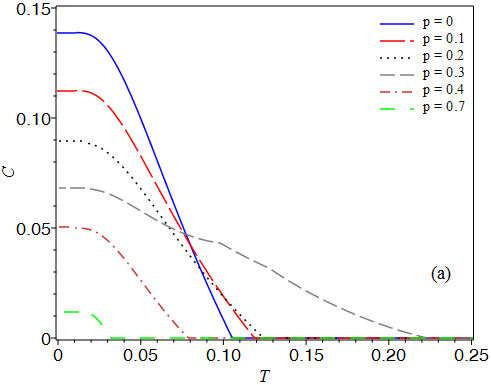}
		\includegraphics[width=7cm,height=5.5cm]{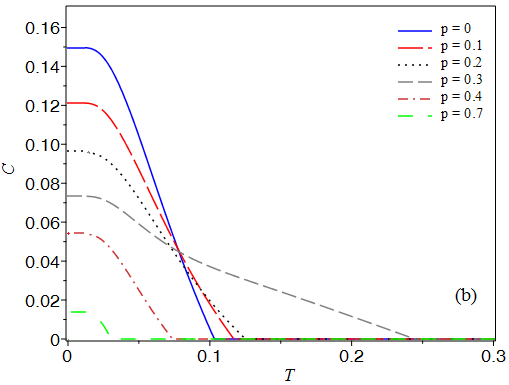}
	 \caption{The concurrence for two half-spins located at the opposite ends of the chain for various fixed values of $p$. (a) When the single boson is fixed at the middle of chain, and (b) when the boson is fixed at the left hand side of the chain. $p>0$ denotes the concurrence under the phase flip. Hamiltonian Parameters have been taken as $J_{1}=-1$, $J_{2}=-0.5$, $\Delta_{1}=-0.5$, $\Delta_{2}=-0.2$ and $B=0.5$.} \label{ConPhaseFlipCent}
	\end{center}
\end{figure}

First we turn our attention to investigate the thermal entanglement between two spins located at the ends of the fermionic chain when the single boson is set at the middle of chain (Fig. \ref{Fig-CentBoson}) with Hamiltonian (\ref{hamiltonian}).
For the mentioned spin formation, we investigate the concurrence at low temperature and weak magnetic field  under decoherence for two half-spins at two ends of the chain and compare it with the case where the phase damping rate is zero (Fig. \ref{ConPhaseFlipCent}).
By inspecting Fig. \ref{ConPhaseFlipCent}(a), one can see that the concurrence gradually decreases upon rising the temperature. As $p$ increases, the maximum amount of concurrence decreases at low temperature. Also, it is obvious that by increasing $p$ until $p \approx 0.3$,  slope of the curve is reduced and the concurrence vanishing goes further on the temperature axis, this actually means that the entanglement can stay alive for higher temperatures. When $p$ increases more, the slope of the concurrence curve increases and the bipartite entanglement vanishes at lower temperatures. 

According to our investigations, the concurrence shows different behaviors with respect to the various positions of the boson. Namely, the single boson movement through the chain can alter the thermal entanglement property between two considered half-spins. For instance, suppose that the boson to be the nearest neighbor of first site of the chain (sub-model 2).

\begin{figure}
	\begin{center}
		\includegraphics[width=7cm,height=5.5cm]{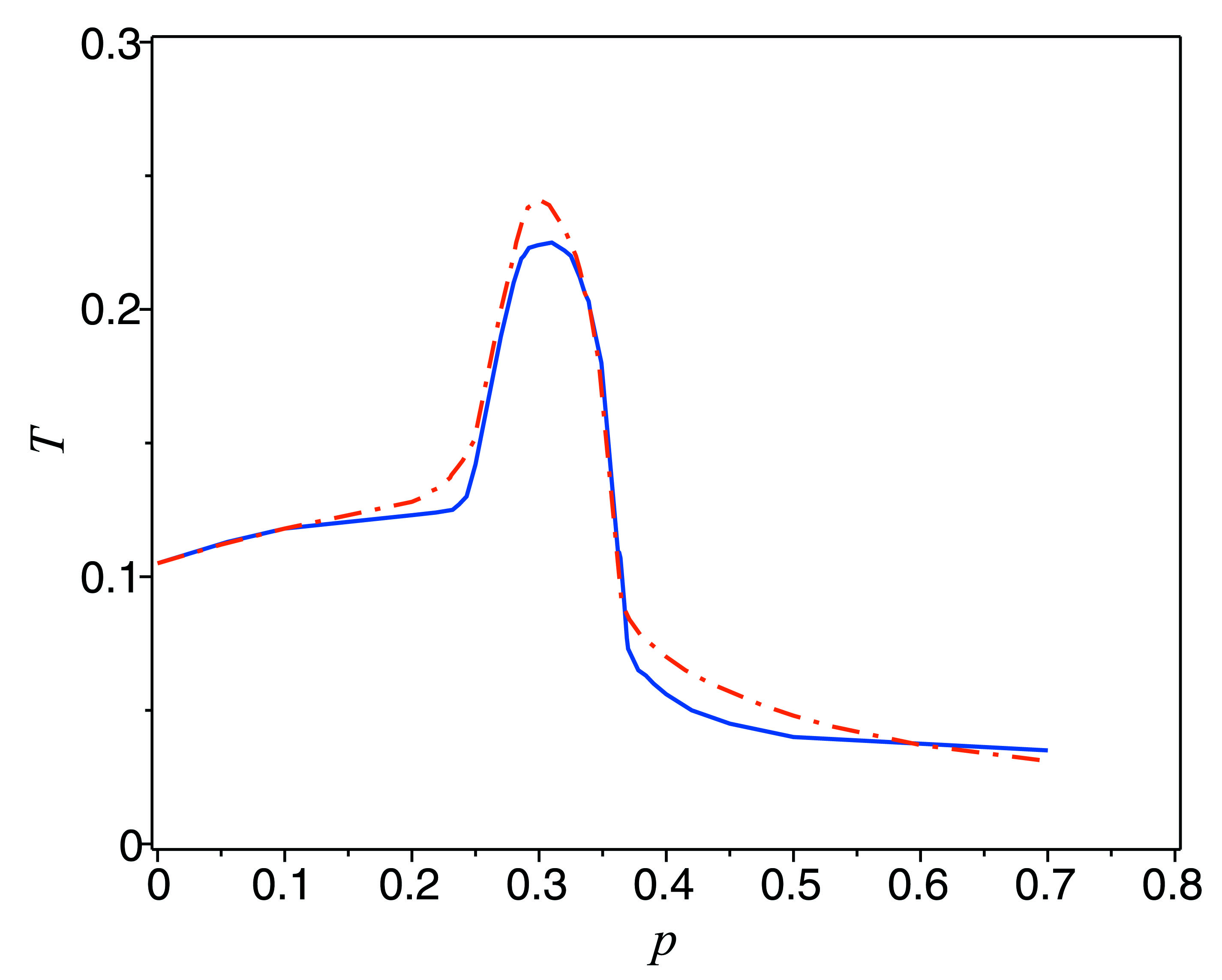}
	\caption{The threshold temperature against decoherence rate $p$ when the boson is situated at the middle of chain (blue solid line) in comparison with that of the boson is placed in the left hand side of the chain (red dashed-dot line) under phase flip.} \label{CritTempVSP}
		\end{center} 
\end{figure}

The concurrence of this model is presented in Fig. \ref{ConPhaseFlipCent}(b), one can see that the maximum value of the concurrence has increased in comparison with Fig. \ref{ConPhaseFlipCent}(a) for all fixed values of $p$. On the other hand, the concurrence vanishing occurs at higher or equal temperatures. We call these temperatures at which the concurrence becomes zero as threshold temperatures. The threshold temperature versus $p$ is illustrated in Fig. \ref{CritTempVSP}, revealing that the optimum point for which the entanglement survives the longest is about $p=0.3$. We understood that the threshold temperature values experience a special sweep point under the considered circumstances.
 It also shows that changing the position of boson to left hand side of the chain (sub-model 2) causes the threshold temperature slightly increase (red dashed-dot line). As a matter of fact, the threshold temperature diagram when the boson is situated at the middle of chain (blue-solid line) can be considered as the threshold temperature boundary.

\begin{figure}
	\begin{center}
		\begin{tikzpicture}
		\draw[-] (0,2) -- (1,2) node[above] {\large{$J_{2}$}};
		\draw[-] (1,2) -- (2,2);
		\draw[-] (2,2) -- (3,2) node[above] {\large{$J_{2}$}};
		\draw[-] (3,2) -- (4,2);
		\draw[-] (4,2) -- (5,2) node[above] {\large{$J_{1}$}};
		\draw[-] (5,2) -- (6,2);
		\draw[-] (6,2) -- (7,2) node[above] {\large{$J_{1}$}};
		\draw[-] (7,2) -- (8,2);
		\shade[ball color=red] (0,2) circle (2ex);
		\shade[ball color=white] (2,2) circle (3.5ex);
		\shade[ball color=red] (4,2) circle (2ex);
		\shade[ball color=red] (6,2) circle (2ex);
		\shade[ball color=red] (8,2) circle (2ex);
		\end{tikzpicture}
		
		\begin{tikzpicture}
		\draw[-] (0,0) -- (1,0) node[above] {\large{$J_{1}$}};
		\draw[-] (1,0) -- (2,0);
		\draw[-] (2,0) -- (3,0) node[above] {\large{$J_{1}$}};
		\draw[-] (3,0) -- (4,0);
		\draw[-] (4,0) -- (5,0) node[above] {\large{$J_{2}$}};
		\draw[-] (5,0) -- (6,0);
		\draw[-] (6,0) -- (7,0) node[above] {\large{$J_{2}$}};
		\draw[-] (7,0) -- (8,0);
		\shade[ball color=red] (0,0) circle (2ex);
		\shade[ball color=red] (2,0) circle (2ex);
		\shade[ball color=red] (4,0) circle (2ex);
		\shade[ball color=white] (6,0) circle (3.5ex);
		\shade[ball color=red] (8,0) circle (2ex);
		\end{tikzpicture}
	 \caption{Top model shows the spin chain containing a single boson located in the left side with spin configuration (1/2,1,1/2,1/2,1/2) (sub-model 2).
		Bottom model shows the spin chain consisting of a single boson localized in the right side with spin configuration (1/2,1/2,1/2,1,1/2) (sub-model 3).} \label{LeftRightBoson}
		\end{center}
\end{figure}
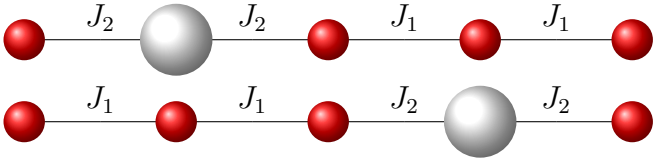
Furthermore, we examine the concurrence for the chain under consideration in which the single boson is the nearest neighbor of the last half-spin (sub-model 3 presented in Fig. \ref{LeftRightBoson}). In this case, we realize that there is no quantum entanglement between two fermions located at the opposite ends of the chain under the considered circumstances, namely, the concurrence is zero. Moreover, we witness that the pairwise entanglement remains zero for a such spin configuration under the phase flip. Finally, we conclude that the position of the single boson in the chain plays an important role to describe the pairwise entanglement intensity between the first and the last half-spins.

\subsection{Bit flip channel}
\begin{figure}[h!]
	\begin{center}
		\includegraphics[width=7cm,height=5.5cm]{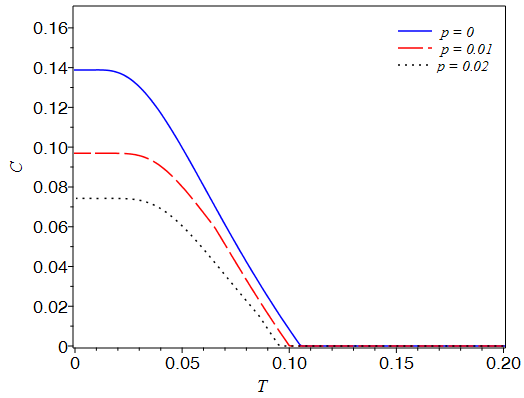}
	 \caption{The concurrence of half-spins at opposite ends of the chain  versus temperature for various fixed values of $p$ under bit flip. The boson is localized at the middle of chain for the same set of the coupling constants as used in Fig. \ref{ConPhaseFlipCent}.}\label{QICB1TBitflip}
		\end{center}
\end{figure}

Now, we investigate three different types of the suggested model under bit flip. The Kraus operators for the bit flip channel are $\Gamma_{{A}}^0=diag(\sqrt{1-p/2},\sqrt{1-p/2})\otimes I_B$, $\Gamma_{{A}}^1=\sqrt{p/2}\sigma^{x}_A\otimes I_B$, $\Gamma_{{B}}^0=I_A\otimes diag(\sqrt{1-p/2},\sqrt{1-p/2})$ and $\Gamma_{B}^1=I_A \otimes \sqrt{p/2}\sigma^{x}_B$ that describe another kind of noise channels between sites ${A}$ and ${B}$. With similar calculations, we obtain the concurrence of state (\ref{density matrices}) under the bit flip as well as the phase flip case. For the case where the single boson is localized at the middle of chain, in Fig. \ref{QICB1TBitflip} we display the temperature dependence of the thermal entanglement between two fermions at the opposite ends of the chain by means of the concurrence. 
The curves indicate the concurrence behavior with respect to the various fixed values of $p$. We see that the thermal pairwise entanglement is very sensitive upon increasing the decoherence rate $p$. When $p$ increases from zero, the maximum value of the concurrence decreases, until for the range  $p \gtrsim 0.05$ the concurrence disappears at low temperature and weak magnetic field.
\begin{figure}[h!]
	\begin{center}
		\includegraphics[width=7cm,height=5.5cm]{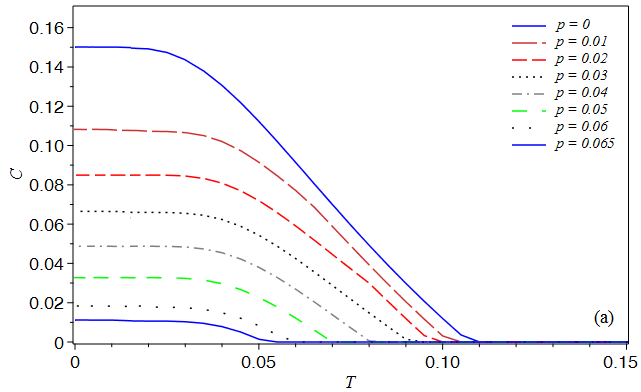}
		\includegraphics[width=7cm,height=5.5cm]{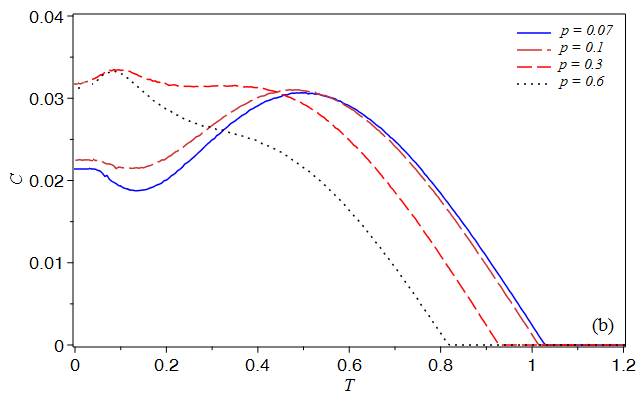}
	 \caption{The concurrence of two half-spins at opposite ends of the chain when the boson is localized at the left side under bit flip (a) in the interval $0\leq p<0.07$, and (b) for $p\geq 0.07$.}\label{QICB3TBitflip}
		\end{center}
\end{figure}

\begin{figure}[h!]
	\begin{center}
		\includegraphics[width=7cm,height=5.5cm]{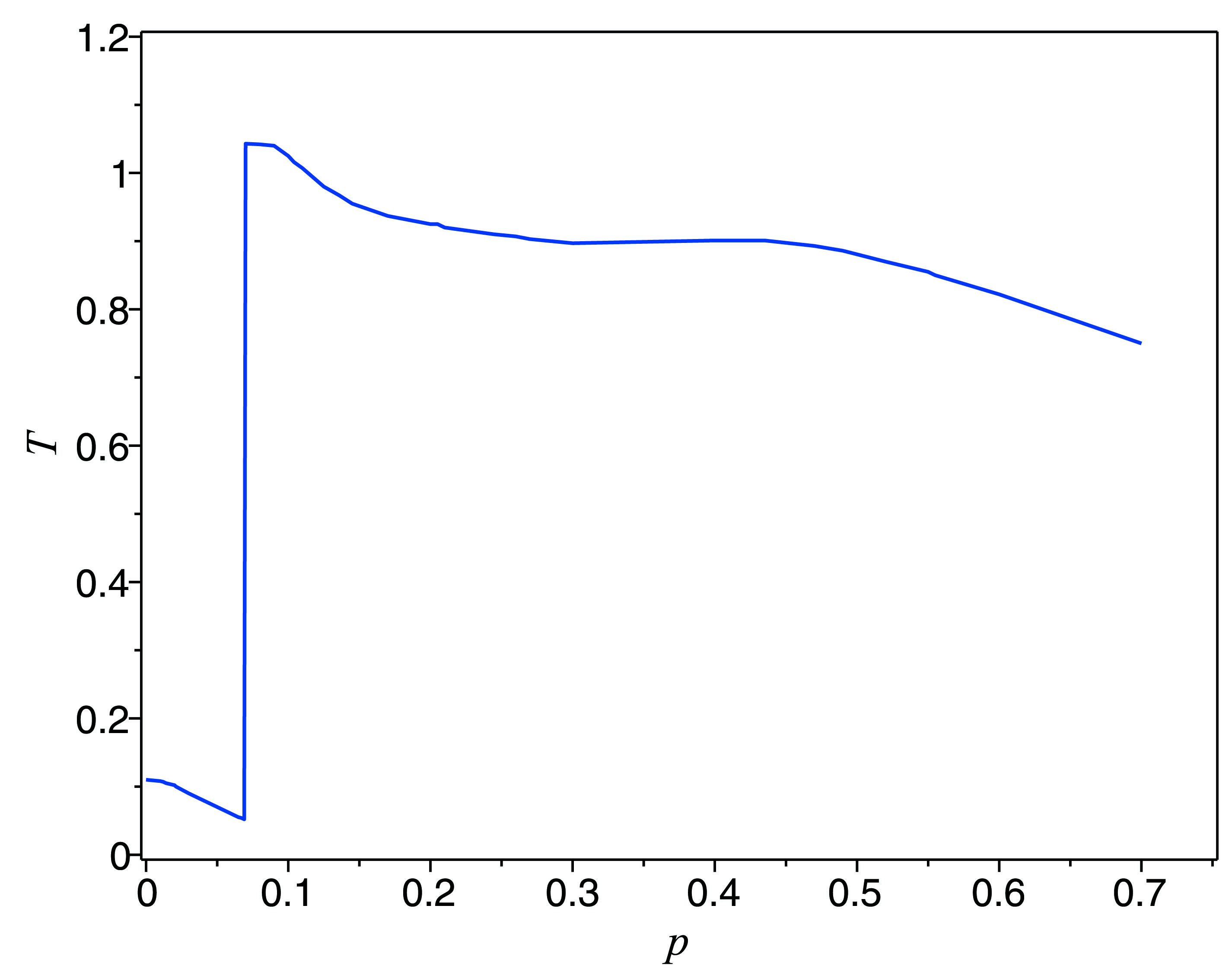}
		\caption{The threshold temperature against decoherence rate $p$ when the boson is the nearest neighbor of the first fermion under bit flip.}\label{Temperature3}
	\end{center}
\end{figure}

In Fig. \ref{QICB3TBitflip} we display the concurrence as a function of temperature for various fixed values of  $p$ for the case where the single boson is boosted  to the left side of the chain. To more clarity, we investigate the behavior of concurrence versus two intervals of $p$, i.e.,
 $0\leq p<0.07$ and $p \ge 0.07$ separately. The maximum amount of concurrence remarkably decreases upon increasing $p$ for the considered interval $0\leq p<0.07$ (Fig. \ref{QICB3TBitflip} (a)), further increase of $p$ (Fig. \ref{QICB3TBitflip} (b)) leads to sharply increasing the concurrence. Also it is obvious that for $p \ge 0.07$, the concurrence continues to exist for higher temperatures. So the entanglement vanishes at higher temperatures. To study this case more accurately, we plot the threshold temperature against decoherence rate $p$ in Fig. \ref{Temperature3}.
In the interval $0 \leq p < 0.07$ the maximum value of concurrence decreases monotonically until reaches a recursive point ($C\approx 0.01$) at which the maximum value starts its increasing with further increase of $p$.  Moreover, in this interval of the decoherence rate, with increase of $p$ the threshold temperature tends to the lower values  until reaches a recursive point at $T\approx 0.05$. By more increasing the decoherence rate $p$ ($p \ge 0.07$), the maximum amount of concurrence shifts toward higher values, on the other hand, the threshold temperature shifts toward higher temperatures as well.
 There is a steep increase of the temperature at $p=0.07$. With further increase of $p$ the threshold temperature begins to decrease again. As a result, the threshold temperature is strongly dependent on the decoherence rate $p$.

It is worthy to note that the concurrence is zero under both phase flip and bit flip for the case where the boson is localized in the right side of the chain at low temperature and weak magnetic field.

\section{Conclusions}\label{Conclusions}
We considered three variants of the antiferromagnetic half-spin chains including a single boson in three different positions (nearest neighbor of the first site, middle of the chain and nearest neighbor of the last site) under decoherence at low temperature and weak magnetic field. We rigorously investigated the thermal entanglement between two fermions at the opposite ends of the chain under the both phase flip and bit flip by means of the concurrence. We realized that the localization of the boson plays an essential role in determining bipartite entanglement intensity in the chain, i.e., by moving the boson along the chain the maximum concurrence remarkably changes. We compared pairwise entanglement of two fermions at the opposite ends of the chain when the single boson is localized at the middle of the chain with the case where the boson is the nearest neighbor of the first fermion and concluded that the maximum value of the concurrence will increase at low temperature. Interestingly in this manner, the entanglement survives for higher temperatures. Furthermore, when the boson is nearest neighbor of the last fermion, the thermal entanglement vanishes under both bit flip and phase flip.

Also, the pairwise thermal entanglement is strongly dependent on the decoherence rate for the various spin arrangements under the phase flip. Indeed, the entanglement death occurs at higher temperatures upon increasing decoherence rate until a threshold point at which the entanglement death turns toward lower temperatures. We investigated the threshold temperature versus decoherence rate and revealed that the threshold temperature increases by increasing the decoherence rate. The maximum amount of the threshold temperature occurs at a specific decoherence rate. With further increasing decoherence rate, the value of the threshold temperature decreases. Here, we proved that the threshold temperature diagram versus the decoherence rate for the case when the boson is localized at the middle of the chain can be considered as the minimum boundary limit of the threshold temperature for all spin configurations of the chain.

Under the bit flip, we concluded that the concurrence decreases upon increasing both temperature and decoherence rate at weak magnetic field. The same as phase flip, the concurrence is dependent on the decohrence rate. We observed a monotonically decreasing behavior of the threshold temperature until a jumping point at which the threshold temperature suddenly increases to its maximum value. After that, the threshold temperature decreases with a gentle slope.

The method introduced in this paper can be straightforwardly generalized to larger quantum spin systems containing different spin arrangements and seeking new interesting outcomes. However, all results should be compatible with the remarks and outcomes expressed in this paper.


\end{document}